\documentclass[twocolumn,showpacs,amsmath,amssymb]{revtex4}

\usepackage{graphics}
\usepackage{bm}

\begin{document}
\title{Linear Responses in Time-dependent Hartree-Fock-Bogoliubov 
Method with Gogny Interaction}
\author{Yukio Hashimoto\\
\it Graduate School of Pure and Applied Sciences, \\
University of Tsukuba, Tsukuba 305-8571, Japan}
\date{\today}
\begin{abstract}
A numerical method to integrate the time-dependent Hartree-Fock Bogoliubov (TDHFB) 
equations with Gogny interaction is proposed. 
The feasibility of the TDHFB code 
is illustrated by the conservation of the energy, particle numbers, 
and center-of-mass in the small amplitude vibrations of ${}^{20}$O.  
The TDHFB code is applied to the isoscalar quadrupole and/or isovector dipole 
vibrations in the linear (small amplitude) region 
in oxygen isotopes ${}^{18,20,22,24}$O, titanium isotopes ${}^{44,50,52,54}$Ti, 
neon isotope ${}^{26}$Ne, and magnesium isotopes ${}^{24,34}$Mg.  
The isoscalar quadrupole and isovector dipole strength functions 
are calculated from the expectation values of the isoscalar quadrupole 
and isovector dipole moments. 
\end{abstract} 
\maketitle
\section{Introduction}
\label{intro}
In nuclear physics,
the mean-field theory has played a central role  
in understanding the nuclear ground state properties 
as well as the dynamical nuclear excitations not only with 
small amplitude but larger one. 
Amomg the microscopic methods in the nuclear mean-field theory,  
the time-dependent Hartree-Fock (TDHF) is characterized 
by the full self-consistency between the mean-field (collective) motion and 
the internal nucleonic motion~\cite{RS}.

The first practical numerical calculation of the TDHF was carried out 
more than thirty years ago~\cite{BKN}. 
Since that time, the TDHF has been applied to the heavy-ion reaction processes 
such as fusion, deep-inelastic collision, fission and so on  
with the purpose of understanding the microscopic mechanism 
of the large amplitude nuclear collective motions in the collision processes~\cite{Negele,Maruhn,Umar}. 
In the TDHF calculations, the three-dimensional (3D) Cartesian grid 
has been used, combined mainly with the Skyrme effective interactions.
Recent TDHF calculations of the reactions of the heavier nuclei 
are carried out by taking advantage of the rapid increase 
of the computer power~\cite{Umar2,Iwata,Kedziora}.

The TDHF has been widely used also in the small amplitude domain  
of the collective motions such as giant resonances (GR). 
It is well known that the TDHF is connected with 
the random phase approximation (RPA) in the small amplitude limit, 
which is a useful tool to study the collective excitations of the nucleus.
Combined with the linear response theory, the excitation energies 
of the collective motions are obtained by calculating the small amplitude responses 
of the nucleus to the impulse type perturbation as an initial condition 
of the TDHF equations~\cite{LinRes}. 
A new method to calculate the RPA energies and amplitudes was proposed 
and applied to several nuclei~\cite{FAM1,FAM2}.

The quasiparticle RPA (QRPA) is an extention of the RPA to include 
the pairing correlation effects. 
In this decade, with the purpose of studying the 
excitation mechanism of the unstable neutron rich nuclei, 
the QRPA calculations have been carried out based upon the various 
formulations
\cite{Matsuo,HagiSaga,Bender02,Khan,Giambrone,Yamagami,Peru05,Terasaki,Bennaceur,Stoitsov,Peru07,Yoshida_C77,Yoshida_C78,Peru08,Tovianen,Losa}.

Just as in the case of the RPA as a small amplitude limit of the TDHF, 
the QRPA is the small amplitude limit 
of the time-dependent Hartree-Fock Bogoliubov (TDHFB). 
The TDHFB is an extension of the TDHF, 
in which the pairing correlation in the nucleus is treated 
selfconsistently with the particle-hole (ph) correlations 
which are taken into account in the TDHF. 

A possibility of the practical numerical calculations 
of the TDHFB with an effective nuclear force (Gogny force) 
was presented in Ref. \cite{YH-KN}. 
The type of full TDHFB calculation was carried out in the 
study of the pairing vibrations with Skyrme interaction~\cite{Avez}. 
Recently, a framework of the TDHFB calculation 
in the canonical form with the Skyrme interaction (Cb-TDHFB) 
was proposed and applied to the linear responses of 
both light and heavy nuclei~\cite{Ebata}. 
The cpu time is substantially reduced even in the calculations of 
the heavier nuclei by using the Cb-TDHFB, which is a useful method 
in studying the systematic property of the 
nulei in the wide range of the nuclear chart.

In the HFB calculations with the Skyrme interactions (Skyrme-HFB), 
the Skyrme interaction is used for the ph channel,
while another type of interaction is introduced 
for the particle-particle (pp) channel.
Since the zero-range interaction is assumed in the Skyrme HFB, 
it is necessary to set the appropriate cut-off energy and choose 
the optimum parameter set in the pairing part \cite{Borycki}. 

In contrast with the Skyrme HFB, in the HFB calculations 
with the Gogny interaction (Gogny-HFB), the ph channel and 
the pp channel are treated on an equal footing. 
In the Gogny-(TD)HFB, a practical cut-off of the energy range of the 
physical space is introduced naturally from the terms 
with the Gaussian functions of the Gogny interaction. 
Therefore, the Gogny interaction is suitable for the formulation 
of the self-consistent TDHFB framework combined with a practical numerical 
method of integrating the TDHFB equations.
We put emphasis on this advantageous point of the Gogny interaction 
and developed a Gogny-TDHFB code. 

In this article, we would like to report the first results of the Gogny-TDHFB 
calculations, illustrating the feasibility of the formulation and numerical codes.
In sect. 2, the basic equations of the TDHFB are simply given. 
A method of solving the TDHFB equations is described.  
In sect. 3, taking oxygen isotope ${}^{20}$O as an example, 
the feasibility of the numerical code of integrating the TDHFB equations 
is illustrated. The conservations of total energy, nucleon numbers, and 
center-of-mass are discussed. 
In sect. 4, the Gogny-TDHFB is applied to 
the linear (small-amplitude) vibrations of spherical nuclei ${}^{18,20,22,24}$O, ${}^{26}$Ne, 
and ${}^{44,50,52,54}$Ti as well as deformed nuclei ${}^{24,34}$Mg as typical examples. 
The strength functions of the isoscalar quadrupole and isovector dipole excitations 
are calculated from the linear responses.
Section 5 is for summary and concluding remarks.
\section{TDHFB equation} \label{sec:1}
The nuclear Hamiltonian is assumed to take the form,
\begin{eqnarray}
  H = \sum_{\alpha \beta} T_{\alpha \beta} C_{\alpha}^{\dagger} C_{\beta} 
     + \frac{1}{4} \sum_{\alpha \beta \gamma \delta} 
         {\cal V}_{\alpha \beta \gamma \delta} C_{\alpha}^{\dagger} C_{\beta}^{\dagger}  
                     C_{\delta} C_{\gamma}, \label{origH} 
\end{eqnarray}
where $T_{\alpha \beta}$ is kinetic energy matrix element and 
${\cal V}_{\alpha \beta \gamma \delta}$ is antisymmetrized two-body matrix element.
The operator $C_{\alpha}^{\dagger} (C_{\alpha})$ is a nucleon creation (annihilation) 
operator of a state labelled with $\alpha$. 

The TDHFB equation for the generalized density matrix ${\cal R}$ 
is written in the form~\cite{RS},
\begin{eqnarray}
   i \hbar \dot{{\cal R}} = \left[ {\cal H}, {\cal R} \right], \label{eq_motion}
\end{eqnarray}
with the HFB Hamiltonian ${\cal H}$,  
\begin{eqnarray}
  {\cal H} = \left( 
             \begin{array}{cc}
               h & \Delta \cr
              - \Delta^{*} & - h^{*} 
             \end{array}             
             \right)\, ,  \label{hfbH}
\end{eqnarray}
and the generalized density matrix ${\cal R}$,  
\begin{eqnarray} 
  {\cal R} = \left( 
             \begin{array}{cc}
               \rho & \kappa \cr
              - \kappa^{*} & 1 - \rho^{*} 
             \end{array}             
             \right)\, , \label{GdenstMat}
\end{eqnarray}
where $\rho$ and $\kappa$ are normal density matrix and pairing tensor, respectively, 
\begin{eqnarray}
\rho_{\alpha \beta} = \left( V^{*} V^{T} \right)_{\alpha \beta}, \, \quad  
\kappa_{\alpha \beta} = \left( V^{*} U^{T} \right)_{\alpha \beta}. \label{rhoVV_kapVU} 
\end{eqnarray}
The matrices $U$ and $V$ are the Bogoliubov transformation matrices,  
\begin{eqnarray}
   \beta_{k}^{\dagger} &=& 
    \sum_{\alpha} \left( U_{\alpha k} C_{\alpha}^{\dagger}  
                       + V_{\alpha k} C_{\alpha} \right), \label{Bog-1} \\
   \beta_{k}           &=& 
    \sum_{\alpha} \left( U_{\alpha k}^{*} C_{\alpha}
                       + V_{\alpha k}^{*} C_{\alpha}^{\dagger}   \right),  \label{Bog-2}
\end{eqnarray}           
from the particle operators $C_{\alpha}^{\dagger}$  and $C_{\alpha}$ 
into the quasi-particles $\beta_{k}^{\dagger}$ and $\beta_{k}$.
The notation $U^{T}$ ($V^{T}$) stands for the transposed matrix of $U$ ($V$). 
The mean field Hamiltonian $h$ and the pairing mean field $\Delta$ are 
introduced through the relations, 
\begin{eqnarray}
  h_{\alpha \beta} &=& T_{\alpha \beta} + \Gamma_{\alpha \beta}, \\ 
 \Gamma_{\alpha \beta} &=& \sum_{\gamma \delta} 
      {\cal V}_{\alpha \gamma \beta \delta} \rho_{\delta \gamma}, \quad
 \Delta_{\alpha \beta} = \frac{1}{2}\sum {\cal V}_{\alpha \beta \gamma \delta} \kappa_{\gamma \delta},
     \label{GamDel}
\end{eqnarray}
with the normal density $\rho$ and the pairing tensor $\kappa$. 

Instead of the equation (\ref{eq_motion}) of the generalized density matrix ${\cal R}$, 
we write the equation of motion in the form~\cite{YH-KN,Bul} 
\begin{eqnarray}
  i \hbar \frac{\partial}{\partial t} 
    \left(
    \begin{array}{c}
                U(t) \cr
                V(t)
             \end{array}
             \right)
     = {\cal H} \left( 
    \begin{array}{c}
                U(t) \cr
                V(t)
             \end{array}
             \right), \label{hfbeq_UV}
\end{eqnarray}
for the matrices $U$ and $V$ in the Bogoliubov transformation (\ref{Bog-1}) and (\ref{Bog-2}).  

Noting that the form of the equation (\ref{hfbeq_UV}) is analogous to 
the TDHF equation~\cite{FCW}, 
we follow the numerical method of integrating the TDHF equation  
and get the formal solution of the TDHFB equation (\ref{hfbeq_UV}), 
\begin{eqnarray}
        \left(
        \begin{array}{c}
                U \cr
                V
             \end{array}
             \right)^{(n+1)} = \exp \left( -i \frac{\Delta t}{\hbar} {\cal H}^{(n+1/2)} \right) 
    \left(
    \begin{array}{c}
                U \cr
                V
             \end{array}
             \right)^{(n)}, \label{tdhfb-solution}
\end{eqnarray} 
where 
$\left( \begin{array}{c} U \cr
                          V
        \end{array} \right)^{(n)}$
is a matrix composed of matrices $U$ and $V$ at a discretized time $t_{n}$ ($n$ = 0, 1, 2, $\cdots$).  
The quantity ${\cal H}^{(n+1/2)}$ is an adequate TDHFB Hamiltonian matrix  
at every time step from $t_{n}$ to $t_{n+1} = t_{n} + \Delta t$.
The Hamiltonian matrix ${\cal H}^{(n+1/2)}$ is determined so that the expectation value $E$ 
of the Hamiltonian (\ref{origH}) with respect to the time-dependent HFB state $| \Phi \rangle$,  
\begin{eqnarray}
  E &=& \langle \Phi | H | \Phi \rangle \nonumber \\
    &=& \sum_{\alpha \beta} T_{\alpha \beta} \rho_{\beta \alpha} 
   + \frac{1}{2} \Gamma_{\alpha \beta} \rho_{\beta \alpha} 
   + \frac{1}{2} \kappa_{\alpha \beta}^{*} \Delta_{\alpha \beta}, \label{Ehfb}
\end{eqnarray} 
is kept constant in the course of the numerical integrations in (\ref{tdhfb-solution}).

Since in the TDHFB calculation the energy conservation is one 
of the most important conditions to be fulfilled, let us see how the 
energy in (\ref{Ehfb}) is conserved with respect to a small variation 
in the matrices $U$ and $V$. 

Let us assume the time step $\Delta t$ is small enough 
to approximate the right hand side of Eq. (\ref{tdhfb-solution}) within  
the first order of $\Delta t$ in the power series expansion of the 
exponential function.  
Putting a parameter $\lambda \equiv \frac{\Delta t}{\hbar}$, 
the variations of the matrices $U$ and $V$ at a time $t$ 
into the new matrices $U'$ and $V'$ are written within the 
first order of the parameter $\lambda$,  
\begin{eqnarray}
 \left( \begin{array}{c}
         U' \\
         V'  \end{array}
 \right)
&=&  \left( \begin{array}{c}
         U \\
         V  \end{array}
 \right)
 - i \lambda \left( \begin{array}{cc}
                    h & \Delta \\
                  - \Delta^{*} & - h^{*} 
                   \end{array}
             \right) 
      \left( \begin{array}{c}
         U \\
         V  \end{array}
      \right) \nonumber \\
 &=& \left( \begin{array}{c}
        U - i \lambda \left( h U + \Delta V \right) \\
        V + i \lambda \left( h^{*} V + \Delta^{*} U \right) 
            \end{array} \right).     \label{UVvariation}
\end{eqnarray}
Here, for the ease of the discussion, 
let us assume also that the time increment $\Delta t$ is small enough 
so that we can identify the mean field Hamiltonian $h^{(n+1/2)}$ 
and mean pairing potential $\Delta^{(n+1/2)}$ in the TDHFB 
Hamiltonian ${\cal H}^{(n+1/2)}$ with $h$ and $\Delta$ at the time $t$, respectively. 

Using the relations in Eq. (\ref{UVvariation}), the variations in the 
density matrix $\rho$ and pairing tensor $\kappa$ in Eq. (\ref{rhoVV_kapVU}) 
are expressed, respectively, as 
\begin{eqnarray}
  & & \rho' = V'^{*} V'^{T} \nonumber \\ 
        &=& \left( V^{*} -i \lambda \left( \Delta U^{*} + h V^{*} \right)  \right)
            \left( V^{T} +i \lambda \left( U^{T} \Delta^{\dagger} + V^{T} h^{\dagger} \right)  \right) \nonumber \\
        &=& \rho -i \lambda [ h, \rho ] - i \lambda \left( - \Delta \kappa^{*} + \kappa \Delta^{*}  \right), 
                                                                                   \label{varrho}  \\
  & & \kappa' = V'^{*} U'^{T} \nonumber \\
          &=& \left( V^{*} -i \lambda \left( \Delta U^{*} + h V^{*} \right)  \right)
            \left( U^{T} -i \lambda \left( U^{T} h^{T} + V^{T} \Delta^{T} \right)  \right) \nonumber \\
        &=& \kappa -i \lambda \left( \Delta U^{*} U^{T} + h \kappa + \kappa h^{*} - \rho \Delta  \right). \label{varkappa}
\end{eqnarray}

We see that the conservation of the particle number in every time step holds 
from Eq. (\ref{varrho}): Taking the trace of both sides in Eq. (\ref{varrho}), we have 
\begin{eqnarray}
   & & {\rm Tr} \rho' - {\rm Tr}\rho \nonumber \\
     &=& - i \lambda \left( {\rm Tr}\left( [h, \rho] \right)
                            + {\rm Tr} \left( - \Delta \kappa^{*} 
                            + \kappa \Delta^{*} \right)  \right)  \nonumber \\
      &=& - i \frac{\lambda}{2} \sum_{\alpha \beta} \sum_{\mu \nu} 
             \left( - {\cal V}_{\alpha \beta \mu \nu} \kappa_{\mu \nu} \kappa^{*}_{\beta \alpha} 
                    + \kappa_{\alpha \beta} \left( {\cal V}_{\beta \alpha \mu \nu} \kappa_{\mu \nu} \right)^{*} \right) \nonumber \\
      & =& - i \frac{\lambda}{2} \sum_{\alpha \beta} \sum_{\mu \nu} 
             \left(  {\cal V}_{\alpha \beta \mu \nu} \kappa_{\mu \nu} \kappa^{*}_{\alpha \beta} 
            - \kappa_{\alpha \beta}  {\cal V}_{\mu \nu \alpha \beta} \kappa_{\mu \nu}^{*}  \right) \nonumber \\
     &=& 0,  \label{num-conserv}
\end{eqnarray}
where the definition of $\Delta$ in Eq. (\ref{GamDel}) and the anti-symmetric property of the matrix elements 
${\cal V}_{\alpha \beta \mu \nu}$ and pairing tensor $\kappa_{\alpha \beta}$ are used. 
The notation Tr stands for taking the trace of the matrices. 
Then, the particle number $N_{\tau} = {\rm Tr} \rho$ is conserved from step to step. 
Here, $N_{\tau}$ is the proton (neutron) number Z (N) for proton (neutron) density 
matrix $\rho = \rho_{p}$ $(\rho_{n}$), respectively. 
\par
Putting the expressions of the density and pairing tensor up to the 
first order in the parameter $\lambda$ in Eqs. (\ref{varrho}) and (\ref{varkappa}) 
into the expression of the energy in Eq. (\ref{Ehfb}),  
we have the variation of the energy as follows:
\begin{eqnarray}
  & & \delta E = {\rm Tr} \bigl\{ -i \lambda h [h, \rho] 
          - i \lambda h \left( - \Delta \kappa^{*} + \kappa \Delta^{*} \right) \bigr\} \nonumber \\
        & & - \frac{i \lambda}{2} 
           {\rm Tr} \bigl\{ - \Delta^{*} \Delta - \Delta^{*} h \kappa - \Delta^{*} \kappa h^{*} 
                            + \Delta^{*} \Delta \rho^{*} + \Delta^{*} \rho \Delta \bigr\} \nonumber \\
        & & - \frac{i \lambda}{2} 
           {\rm Tr} \bigl\{   \quad \Delta^{*} \Delta + h^{*} \kappa^{*} \Delta + \kappa^{*} h \Delta 
                            - \Delta^{*} \rho \Delta - \rho^{*} \Delta^{*} \Delta \bigr\} \nonumber \\
        & & = 0, \label{e-conserv}
\end{eqnarray}
where the relations $h^{*} = h^{T}$, $\kappa^{T} = - \kappa$, 
and $\Delta^{T} = - \Delta$ are used. 
From Eq. (\ref{e-conserv}), we see that 
we can integrate the TDHFB equation (\ref{hfbeq_UV}), conserving the energy 
expressed as in Eq. (\ref{Ehfb}), by appropriately setting the time increment 
$\Delta t$ and intermediate TDHFB Hamilonian ${\cal H}^{(n+1/2)}$. 
In the case of the Gogny interaction \cite{Gogny,Gogny2} with the parameter set D1S, 
the contribution of the density-dependent term is 
included only in the mean-field Hamiltonian $h$ 
in the HFB Hamiltonian (\ref{hfbH}), through the variation 
of the nucleon density $\rho({\bf x})$ \cite{Girod}.
Then, the energy conservation relation (\ref{e-conserv}) 
holds when the Gogny interaction with the parameter set D1S 
is adopted in the Hamiltonian (\ref{origH}).

\section{Numerical solution and conserved quantities} \label{sec:3}
\subsection{Parameters} \label{sec:3-1}
With the purpose of demonstrating the feasibility of our TDHFB code, 
we apply them to the small amplitude (linear) responses of the 
oxygen isotope ${}^{20}$O.
We adopt the Gogny interaction \cite{Gogny,Gogny2} for the two-particle interaction part 
in the Hamiltonian (\ref{origH}). 
The parameter set D1S is used. 
The Coulomb part is omitted in all of the calculations in this article.

The initial condition in the present calculation is of the impulse type:
The static HFB solutions $U_0$ and $V_0$ are changed into the initial matrices 
$U^{(0)}$ and $V^{(0)}$ by the relations 
\begin{eqnarray}
  V^{(0)} &=& \exp \left( i \varepsilon {\bf Q} \right) V_0 
           = \sum_{\nu = 1}^{N_{max}} 
           \frac{i^\nu \varepsilon^\nu {\bf Q}^\nu}{\nu!} V_0,  \label{ini-1} \\
  U^{(0)} &=& \exp \left( -i \varepsilon {\bf Q}^{*} \right) U_0  
           = \sum_{\nu = 1}^{N_{max}} 
           \frac{i^\nu (-\varepsilon)^\nu {{\bf Q}^{*}}^\nu}{\nu!} U_0,  \nonumber \\
          & &                                                             \label{ini-2} 
\end{eqnarray}
where ${\bf Q}$ stands for the matrix representation 
of a multipole operator with respect to the numerical basis states. 
$N_{max}$ is the maximum number up to which the exponential function is 
expanded into power series. 
The parameter $\epsilon$ is chosen so that the linearity is held in the 
course of the time integration of the equations (\ref{hfbeq_UV}).

\begin{figure}[htb]
\begin{center}
\resizebox{0.4\textwidth}{!}{%
  \includegraphics{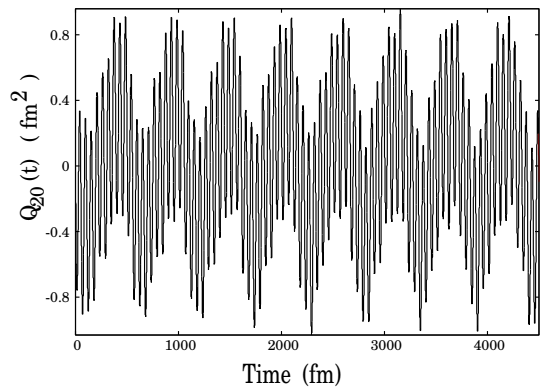}
}
\end{center}
\caption{Time dependence of expectation value of isoscalar quadrupole moment 
of ${}^{20}$O. 
}
\label{quadosci}
\end{figure}

\begin{figure}[htb]
\begin{center}
\resizebox{0.4\textwidth}{!}{%
  \includegraphics{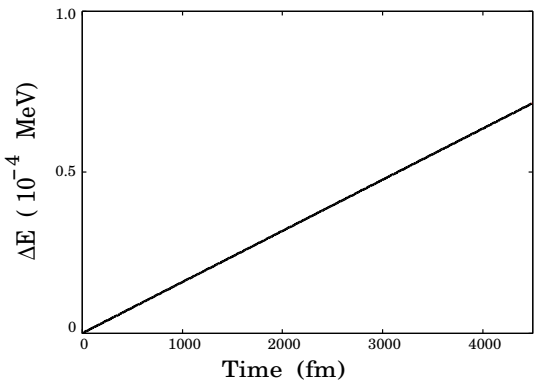}
}
\end{center}
\caption{Deviation of the excitation energy from the initial (t = 0) value 
of the isoscalar quadrupole vibration in fig. \ref{quadosci} with respect to time. }
\label{oscienergy}
\end{figure}

\begin{figure}[htb]
\begin{center}
\resizebox{0.4\textwidth}{!}{%
  \includegraphics{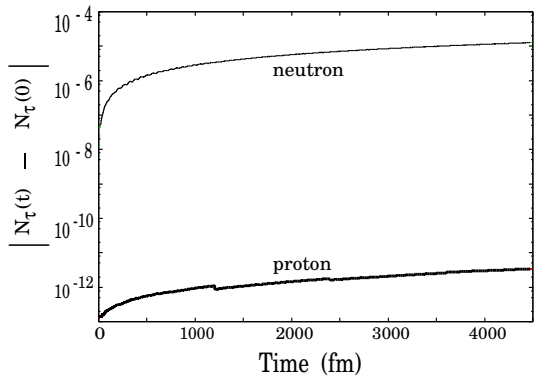}
}
\end{center}
\caption{Deviation of neutron (proton) number expectation value 
from the accurate number 12 (8), respectively, in isoscalar quadrupole vibration 
in fig. \ref{quadosci}.
}
\label{numb_consv}
\end{figure}

In making the TDHFB hamiltonian ${\cal H}$, we used the same form of energy functional 
as that used in the calculation of the HFB ground state, i.e., 
the center-of-mass correction part $E_{CM} = - {\bf P}^2/2Am$ 
with the nucleon mass $m$ and 
the total momentum operator ${\bf P} = \sum_i^{A} {\bf p}_i$ of $A$ nucleons is included. 

As the numerical basis, the three-dimensional harmonic oscillator eigenstates were used. 
The two-particle matrix elements ${\cal V}_{\alpha \beta \gamma \delta}$ 
in the Hamiltonian (\ref{origH}) were calculated by following the method 
described by Girod and Grammaticos \cite{Girod}. 

We used the space of basis states with the relation $0 \leq n_x + n_y + n_z \leq 4$,  
where $n_x$ ($n_y$, $n_z$) is the number of quanta 
of the harmonic oscillator basis states in the x (y, z) direction, respectively. 
The harmonic oscillator parameters were set as follows:  
$\hbar \omega_x = \hbar \omega_y = \hbar \omega_z$ = 13.7 MeV.

In the initial conditions (\ref{ini-1}) and (\ref{ini-2}), 
we took the multipole operator ${\bf Q}$ to be 
an isoscalar quadrupole operator $Q_{\alpha \beta} 
= \left( 2 z^2 - x^2 - y^2 \right)_{\alpha \beta}$.
The parameter $\varepsilon$ was put to be $1.0 \times 10^{-3}$, 
being small enough so that the linearity of the oscillation 
with respect to $\varepsilon$ is satisfied.
The power series expansions of the exponential functions were taken up to 
the $N_{max}$-th order with $N_{max}$ put to be ten.

In the course of the time integrations in (\ref{tdhfb-solution}), 
a predictor-corrector method was made use of 
in making the Hamiltonian matrix ${\cal H}^{(n+1/2)}$ at every time step~\cite{FCW}. 

\subsection{Conservation of energy, particle number, and center-of-mass}
\label{sec:3-2}
In fig. \ref{quadosci}, we display the time variation of the 
isoscalar quadrupole moment 
$Q_{20}(t) = \langle  \sum_{i=1}^{A} r_i^2 Y_{20}(\hat{r}_i)  \rangle$ of ${}^{20}$O. 
The time increment $c \Delta t$ is 0.2 fm. 
After the initial impulse, regular small-amplitude oscillations take place. 

In fig. \ref{oscienergy}, the deviation of the excitation energy 
from the initial (t = 0) value in the course of the vibration is shown. 
The excitation energy varies from the initial value 0.02554 MeV to 
0.02561 MeV after integration time 4500 fm (22500 steps). 
The deviation of the excitation energy from the initial value 
in the course of the time integration is reduced 
when we carry out more numbers of predictor-corrector loops  
in each time step. 
 
In fig. \ref{numb_consv}, the deviation of the expectation values 
of the nucleon number $N_{\tau}$ 
from the accurate values (8 protons and 12 neutrons) 
is displayed with respect to time. 
In ${}^{20}$O, the protons are in the normal state, whereas the 
pairing correlation is active on the neutron side. 
Therefore, the integration of the equations of motion 
of the proton orbitals is equivalent to the 
TDHF case. Then, the total number of protons are conserved within 
$10^{-11}$. 
The neutron number in fig. \ref{numb_consv} is kept 
within around $10^{-5}$ in the present integration process.
These results illustrate that the unitarity of the time-displacement operator 
in Eq. (\ref{tdhfb-solution}) holds within this accuracy. 

\begin{figure}[htb]
\begin{center}
\resizebox{0.45\textwidth}{!}{%
  \includegraphics{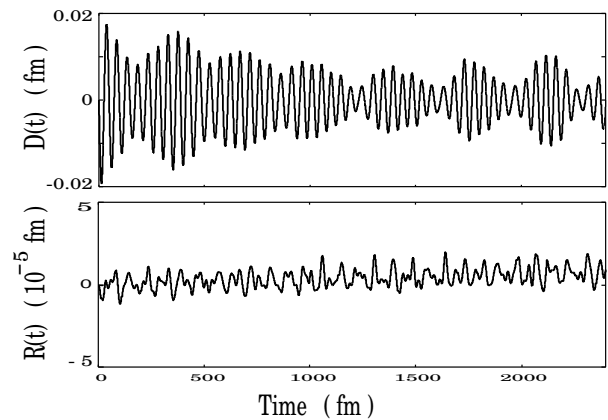}
}
\end{center}
\caption{Time dependence of expectation values of isovector 
dipole moment D(t) (upper panel) and center-of-mass R(t) (lower panel).
}
\label{IVD-oscillation}
\end{figure}

Taking the initial impulse operator $Q_{\alpha \beta}$ to be 
the isovector dipole operator 
$D_{\alpha \beta} = \frac{N}{A} \left( z \right)_{\alpha \beta}$
 for protons and $ - \frac{Z}{A} \left( z \right)_{\alpha \beta} $ for 
neutrons, we have the oscillation pattern in fig. \ref{IVD-oscillation}, 
where the expectation value 
$D(t) = \langle \frac{N}{A} \sum_{i=1}^{Z} z_i - 
\frac{Z}{A} \sum_{i=1}^{N} z_i \rangle $ is displayed together with 
the center-of-mass expectation value in z direction 
$R(t) = \frac{1}{A}\langle \sum_{i = 1}^{Z} z_i + \sum_{i = 1}^{N} z_i \rangle$ 
in the first 12000 steps of integration. 
Here, $N (Z)$ stands for the neutron (proton) number, respectively.  
We see that the fluctuation of the center-of-mass motion $R(t)$ is 
kept within $2 \times 10^{-5}$ fm. 
The effect of the center-of-mass motion on the isovector dipole moment $D(t)$ 
could be neglected.

When the quadrupole moment operator ($2 z^2 - x^2 - y^2$ or $x^2 - y^2$) is used 
as the impulse operator, the expectation values of the center-of-mass and off-diagonal terms 
$xy$, $yz$, and $zx$ are kept to be zero, 
since the symmetry of the system at the initial time $t = 0$ 
is kept throughout the integration time. 
\section{Applications} \label{sec:4}
\subsection{Spherical nuclei} \label{sec:4-1}
\subsubsection{${}^{18,20,22,24}$O } \label{sec:4-1-1}
\begin{figure}[h]
\begin{center}
\resizebox{0.4\textwidth}{!}{%
  \includegraphics{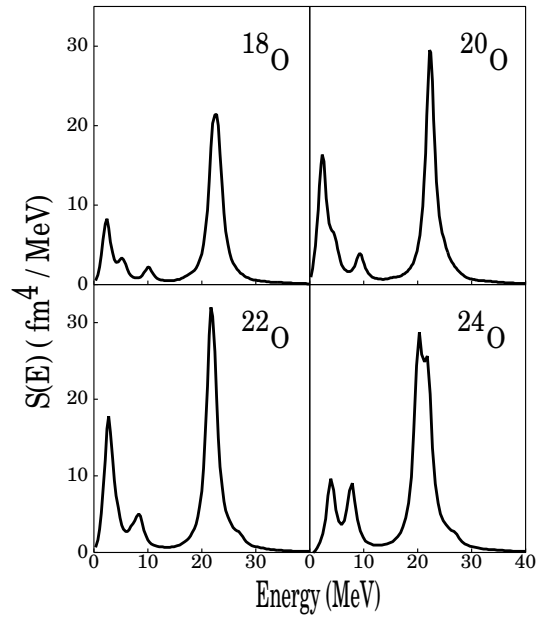}
}
\end{center}
\caption{Strength functions of isoscalar quadrupole mode 
of oxygen isotopes ${}^{18,20,22,24}$O. Artificial width of 1 MeV is used. }
\label{oxygen_strength}
\end{figure}

Using the impulse type initial conditions (Eqs. (\ref{ini-1}) and (\ref{ini-2}))  
with an isoscalar quadrupole operator \\
$Q_{\alpha \beta} = \left( 2 z^2 - x^2 - y^2 \right)_{\alpha \beta}$ 
and the space of basis functions $0 \leq n_x + n_y + n_z \leq 4$, we applied the method 
of integrating the TDHFB equations discussed in the previous section 
to the oxygen isotopes ${}^{18,20,22,24}$O. 
Taking the Fourier transformation of the quadrupole moment $Q_{20}(t)$, 
we calculated the strength functions of the isoscalar quadrupole mode of the oxygen isotopes 
in fig. \ref{oxygen_strength}~\cite{LinRes}.

\begin{figure}[h]
\begin{center}
\resizebox{0.4\textwidth}{!}{%
  \includegraphics{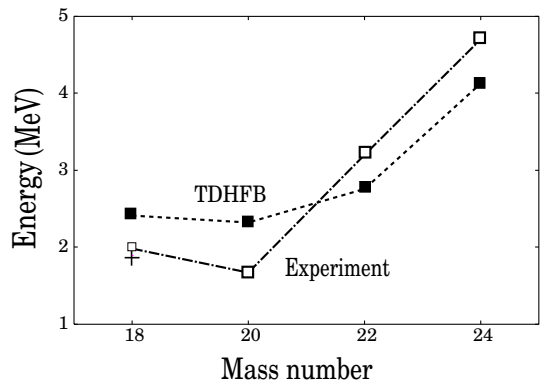}
}
\end{center}
\caption{ Energies of the first peaks of the isoscalar quadrupole strength functions 
of oxygen isotopes ${}^{18,20,22,24}$O in fig. \ref{oxygen_strength}. 
Closed (open) squares stand for the TDHFB (experimental) results, respectively. 
Plus (+) is the first 2$^{+}$ energy of neon ${}^{18}$Ne~\cite{Tilley}.  }
\label{ox-18-24_energies}
\end{figure}

In fig. \ref{oxygen_strength}, we see that the low-energy peaks 
are located below 5 MeV. 
The energies of the peaks below 5 MeV are shown in fig. \ref{ox-18-24_energies}. 
The neutron pairing energies of the oxygen isotopes are shown in table 1. 
In the four oxygen isotopes ${}^{18,20,22,24}$O, the magnitude of the neutron pairing energy 
is the largest in ${}^{20}$O and zero in ${}^{24}$O. 
The neutron number 16 in ${}^{24}$O corresponds to the subshell closure of 2s$\frac{1}{2}$.  
The tendency of the energies of the peaks in the TDHFB calculation that the lowest (highest) is 
of ${}^{20}$O (${}^{24}$O) is in accordance with the experimental data~\cite{AtomData,Frank,Hoffman}. 
In the present TDHFB calculation, the energy of the lowest peak of ${}^{24}$O is 4.1 MeV, 
which is similar to the QRPA predictions around 4 MeV~\cite{Khan,Obertelli}. 
The experimental results of the energies of the lowest 2${}^{+}$ state in ${}^{24}$O are 
around 4.7 MeV \cite{Frank,Hoffman}, 
which is higher than the QRPA and present TDHFB results by around 0.7 MeV.

Because the Coulomb force is omitted in the present calculations, 
we compare our result of the first 2${}^{+}$ energy of the oxygen ${}^{18}$O 
with the experimental value of neon ${}^{18}$Ne. 
The experimental values of the first 2${}^{+}$ energies of the oxygen ${}^{18}$O and neon ${}^{18}$Ne 
are 1.98 MeV and 1.89 MeV~\cite{Tilley}, respectively. 
These two nuclei are mirror nuclei to each other and their first 2${}^{+}$ energies 
are located within the width of 0.1 MeV. 
The present TDHFB result of the oxygen ${}^{18}$O is around 0.5 MeV 
higher than these experimental results.

\begin{figure}[htb]
\begin{center}
\resizebox{0.4\textwidth}{!}{%
  \includegraphics{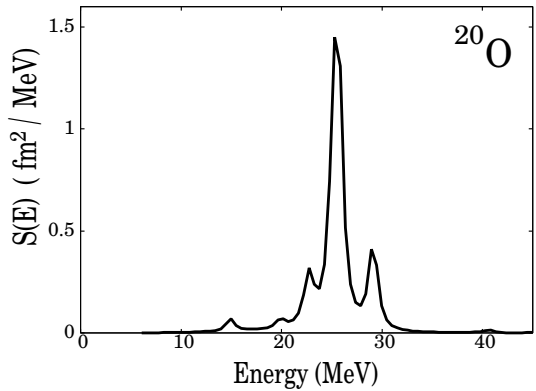}
}
\end{center}
\caption{ Strength function of isovector dipole mode of $^{20}$O. 
Artificial width is 0.5MeV. }
\label{IVD-strength}
\end{figure}

Through the Fourier transformation of the isovector dipole moment $D(t)$ 
of ${}^{20}$O in fig. \ref{IVD-oscillation}, we get the strength function  
of the isovector dipole mode in fig. \ref{IVD-strength}.
The peak is located at 25 MeV, which is larger than the observed value 
by a few MeV \cite{oxygen-isotopes}.  
Since the absolute values of the center-of-mass is kept practically small 
in the time integration, the effect of the spurious motion in the strength function 
might be neglected. 

\begin{table}[h]
\caption{
Neutron pairing energy E$_{\rm n}$ in oxygen isotopes ${}^{16,18,20,22,24}$O 
in Gogny-HFB.
}
\label{table-1}
\begin{center}
\begin{tabular}{c|ccccc}
                      & ${}^{16}$O    & ${}^{18}$O   & ${}^{20}$O     &   ${}^{22}$O  &  ${}^{24}$O     \\ 
\hline 
E$_{\rm n}$ (MeV)     &     0         &    - 4.56    & - 5.33         &   - 2.39      &      0  
\end{tabular}
\end{center}
\end{table}

\subsubsection{Pairing energy in ${}^{20}$O} \label{sec:4-1-2}
\begin{figure}[h]
\begin{center}
\resizebox{0.45\textwidth}{!}{%
  \includegraphics{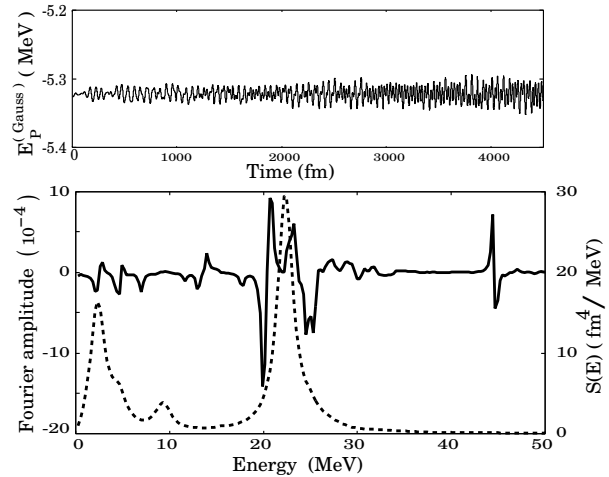}
}
\end{center}
\caption{(Upper panel) Time dependence of the pairing energy $E_{p}^{\rm (Gauss)}$ 
of the Gaussian part in the Gogny interaction in the quadrupole oscillation 
in fig. \ref{quadosci} of ${}^{20}$O. 
(Lower panel) Fourier (sine) amplitude of the fluctuations 
of the pairing energy $E_{p}^{\rm (Gauss)}$ in the upper panel (solid curve, left axis).
Artificial width of 0.25 MeV is used.  
The isoscalar quadrupole strength function is plotted in dashed curve (right axis).}
\label{ox-20_epgss_strength}
\end{figure}

In fig. \ref{ox-20_epgss_strength}, we show the time dependence of the pairing energy 
$E_{p}^{(\rm Gauss)} = \frac{1}{4} \sum_{\alpha \beta \mu \nu} \kappa_{\alpha \beta}^{*} 
{\cal V}_{\alpha \beta \mu \nu}^{(\rm Gauss)} \kappa_{\mu \nu}$ \cite{RS} 
of the Gaussian part ${\cal V}_{\alpha \beta \mu \nu}^{(\rm Gauss)}$ in the Gogny interaction 
in the quadrupole oscillations in fig. \ref{quadosci} of ${}^{20}$O. 
Here, $\kappa_{\mu \nu}$ is the pairing tensor of the neutrons.
The magnitude of the pairing energy which comes from the other parts 
in the Gogny interaction, i.e., 
spin-orbit, center-of-mass, and Coulomb (if included in the 
calculations) terms, is small compared with that from the Gaussian part. 
Since the initial impulse is taken to be small enough to ensure the linearity of the 
quadrupole oscillation, the fluctuation of the pairing energy 
in fig. \ref{ox-20_epgss_strength} is within 0.05 MeV in the integration time up to 
4500 fm. 

We take the Fourier transformation of the fluctuation of the pairing energy 
and show the Fourier (sine) amplitudes in the lower panel 
in fig. \ref{ox-20_epgss_strength}. 
Corresponding to the giant resonance peak in the strength function in the 
energy range from 20 MeV to 25 MeV, we see remarkable fluctuations 
in the Fourier amplitudes. In the low energy region under 7 MeV, we find 
fluctuations in the Fourier amplitudes corrresponding to the low-energy peak
at 2.3 MeV and a shoulder around 4.4 MeV in the strength function. 
On the other hand, there are fluctuations in the Fourier amplitudes to which 
there are no corresponding peaks in the strength function curve. 

In the present formulation of the TDHFB, 
we can not separate out the effects of the pairing correlation, 
since they are embedded in the dynamical changes of the matrices 
U and V which are defined on the fixed basis of 3D harmonic oscillator 
eigenfunctions. We need some theoretical tools to separate out the 
effects of the pairing correlation. 
We expect that the idea of the Cb-TDHFB will be a good starting point 
to improve the present formulation and attack the problem 
of making clear the pairing effects in the dynamical processes 
of nuclear collective vibrations not only of the quadrupole type 
but of the other multipolarity types such as monopole and octupole.

\subsubsection{${}^{44,50,52,54}$Ti} \label{sec:4-1-3}
The next example of the TDHFB calculations in the spherical nuclei 
is the application of the method to the isoscalar quadrupole vibrations 
in the titanium isotopes ${}^{44,50,52,54}$Ti.
Here and hereafter, the space of the basis functions $0 \leq n_x + n_y + n_z \leq 5$ 
was used in the calculations. 

\begin{figure}[h]
\begin{center}
\resizebox{0.4\textwidth}{!}{%
  \includegraphics{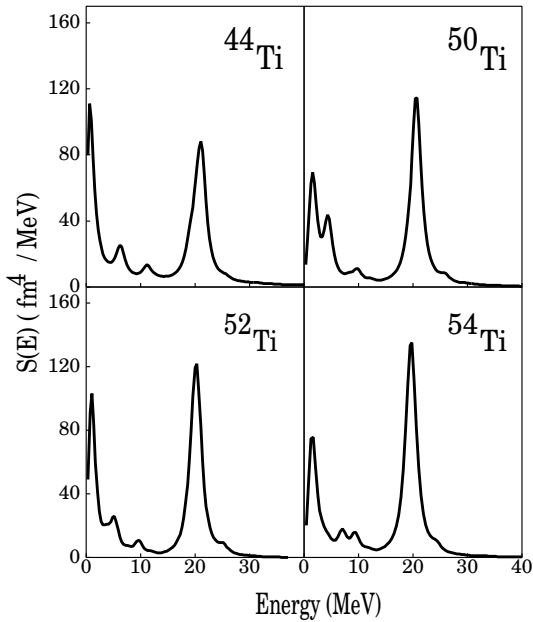}
}
\end{center}
\caption{Strength functions of isoscalar quadrupole mode of titanium isotopes ${}^{44,50,52,54}$Ti.
Artificial width of 1 MeV is used. }
\label{ti-44-54}
\end{figure}

\begin{figure}[h]
\begin{center}
\resizebox{0.4\textwidth}{!}{%
  \includegraphics{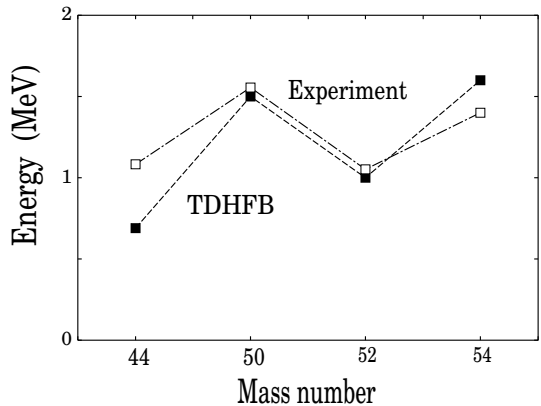}
}
\end{center}
\caption{Energies of the first peaks of the strength functions 
of the isoscalar quadrupole mode of the titanium isotopes ${}^{44,50,52,54}$Ti  
in fig. \ref{ti-44-54}. 
Closed (open) squares stand for the TDHFB (experimental) 
results, respectively~\cite{Simpson,Janssens}. }
\label{ti-energies}
\end{figure}

The strength functions of the titanium isotopes are shown in fig. \ref{ti-44-54}. 
From the strength functions, we got the energies of the first low-energy peaks,  
which are shown in fig. \ref{ti-energies}. 
In fig. \ref{ti-energies}, we see that the TDHFB results of the lowest-energy 2${}^{+}$ peaks are 
along the experimental values~\cite{Simpson,Janssens}: 
The lowest 2${}^{+}$ energies in ${}^{50}$Ti and ${}^{54}$Ti 
are higher than the other two nuclei ${}^{44}$Ti and ${}^{52}$Ti. 

This tendency of the lowest 2${}^{+}$ energies of the four nuclei 
is in accordance with the tendency of the neutron pairing energies 
and sushell closures. In table 2, we show the pairing energies of protons and neutrons 
in the four nuclei ${}^{44,50,52,54}$Ti in their HFB ground states.  
The proton pairing energies in ${}^{44,50,52,54}$Ti are almost constant around - 4.8 MeV. 
In ${}^{50}$Ti, the neutron number is 28 and 1f$\frac{7}{2}$ orbitals are filled. 
The pairing energy in ${}^{50}$Ti is zero. 
Likely, in  ${}^{54}$Ti, 2p$\frac{3}{2}$ orbitals are filled 
and the pairing energy is comparatively small. 

\begin{table}[h]
\caption{
Proton (neutron) pairing energy E$_{\rm p}$ (E$_{\rm n}$) in titanium isotopes ${}^{44,50,52,54}$Ti 
in Gogny-HFB.
}
\label{table-2}
\begin{center}
\begin{tabular}{c|cccc}
                                 & ${}^{44}$Ti    & ${}^{50}$Ti   & ${}^{52}$Ti     &   ${}^{54}$Ti     \\ 
\hline 
E$_{\rm p}$ (MeV)                 & - 4.92         &  - 4.72        &  - 4.79         & - 4.80       \\   
E$_{\rm n}$ (MeV)                 & - 4.92         &    0            & - 2.75          & - 1.17    
\end{tabular}
\end{center}
\end{table}

\subsubsection{${}^{26}$Ne} \label{sec:4-1-4}
The mean-field of the ground state of ${}^{26}$Ne is spherical in the Gogny-HFB 
calclation. The protons are in the paired state 
whereas the neutron pairing energy vanishes. 
${}^{26}$Ne is one of the neutron-rich nuclei, and a pygmy state is known 
as the low-energy components in the isovector dipole resonance~\cite{Gibelin}.   

\begin{figure}[htb]
\begin{center}
\resizebox{0.38\textwidth}{!}{%
  \includegraphics{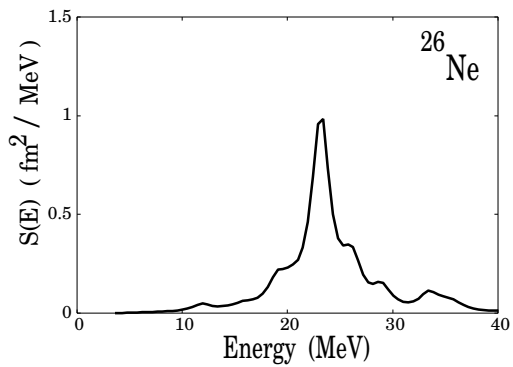}
}
\end{center}
\caption{Strength function of isovector dipole mode of ${}^{26}$Ne.}
\label{ne26-strength}
\end{figure}

Taking the isovector dipole operator as the impulse operator, we calculated the 
expectation values of the isovector dipole moment in ${}^{26}$Ne. 
We show the strength function in fig. \ref{ne26-strength}.

The energy of the small 1${}^{-}$ peak near 10 MeV is 11.9 MeV 
in the present TDHFB calculation. 
In the QRPA calculation by Peru at al.~\cite{Peru07}, 
they got the value 10.64 MeV for the 1${}^{-}$ peak of the pygmy state. 
The experimental value of the pygmy state is around 9 MeV~\cite{Gibelin}. 
The corresponding energy by the present TDHFB calculation is larger than 
the QRPA (experimental) result by 1.3 (3) MeV, respectively. 
 
\subsection{Deformed nuclei} \label{sec:4-2}
As the typical examples of the deformed nuclei, we took magnesium isotopes  ${}^{24}$Mg 
and ${}^{34}$Mg. 
In ${}^{24}$Mg, the pairing energies of both protons and neutrons are zero 
in the HFB ground state, in which the mean-field is in a prolate form. 
${}^{34}$Mg is also prolately deformed in the HFB ground state,  
in which the neutron pairing energy is finite.  
The deformation parameter $\beta$ of ${}^{24}$Mg (${}^{34}$Mg) is 
$\beta$ = 0.51 (0.43), respectively~\cite{HilGirod}.   

\begin{figure}[h]
\begin{center}
\resizebox{0.4\textwidth}{!}{%
  \includegraphics{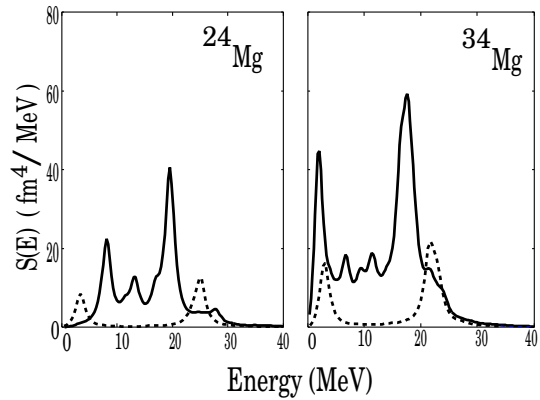}
}
\end{center}
\caption{Strength functions of isoscalar quadrupole mode of magnesium isotopes ${}^{24,34}$Mg.
Solid (dashed) curve stands for $K$ = 0 (2) component, respectively. 
Artificial width of 1 MeV is used.}
\label{mg-24-34}
\end{figure}

Calculating the time dependence of the 
expectation values of the $K$ = 0 and 2 components of the isoscalar quadrupole moment, 
we get the strength functions in fig. \ref{mg-24-34}. 
Here we used the isoscalar quadrupole operator $x^2 - y^2$ as the initial impulse operator 
to excite the $K =$ 2 component of the isoscalar quadrupole vibrations.

The giant resonance peaks with $K = $ 0 and 2 are seen in the energy region 
from 15 MeV to 30 MeV in both of the nuclei ${}^{24}$Mg and ${}^{34}$Mg. 
The peaks with $K =$ 2 are located in the higher energy region than those 
with $K =$ 0, reflecting the prolate deformation of the mean-fields in 
the ground states of the two nuclei.  

The energy of the lowest $K = $ 0 (2) peak in ${}^{24}$Mg is 8.06 (3.3) MeV, 
which is comparable with an experimental value 6.43 (4.23) MeV, respectively~\cite{Endt}. 
The characteristic point that the $K =$ 2 peak comes lower in energy 
than that with $K =$ 0 is closely related with the softness of  ${}^{24}$Mg 
with respect to the direction of the gamma deformation.

In ${^{34}}$Mg, the energy of the lowest $K =$ 0 (2) peak is 
2.07 (3.0) MeV. In the QRPA calculations with the Skyrme interaction, 
the energy of the corresponding peak is 2.65 (around 3.0) MeV~\cite{Yoshida_C78,Losa}.
The energies of the two peaks with $K = $ 0 and 2 by the TDHFB calculations 
are comparable with those by the QRPA calculations, though the former tend to 
be smaller than the latter. 
 
\begin{figure}[h]
\begin{center}
\resizebox{0.4\textwidth}{!}{%
  \includegraphics{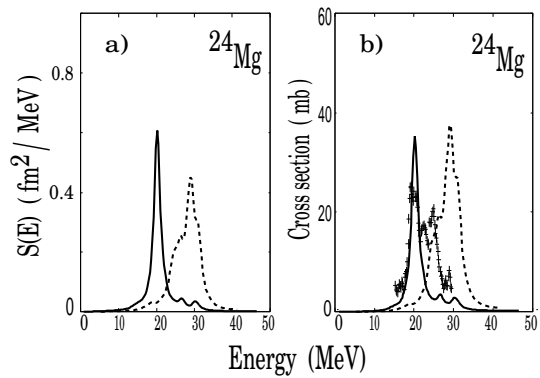}
}
\end{center}
\caption{Strength functions of isovector dipole mode (a)) and 
photoabsorption cross sections (b)) of ${}^{24}$Mg.
Solid (dashed) curve stands for $|K|$ = 0 (1) component, respectively.
Artificial width of 1 MeV is used in the strength functions. 
Experimental data of cross sections are shown with plus (+)~\cite{Varlamov}. }
\label{mg-24-IVD}
\end{figure}

The strength functions of the isovector dipole modes of 
${}^{24}$Mg are shown in the left panel in fig. \ref{mg-24-IVD}. 
The energy of the $K = $ 0 (1) peak in fig. \ref{mg-24-IVD} 
is 20.2 (28.9) MeV, respectively.
The corresponding value obtained by using the QRPA by Peru et al. 
is 19.35 (27.44) MeV~\cite{Peru08}. 
The present TDHFB result of the $K =$ 0 (1) peak energy is higher by 0.9 (1.5) MeV 
than the QRPA result, respectively. 

The strength functions of the isovector dipole modes of 
${}^{24}$Mg are shown in the left panel in fig. \ref{mg-24-IVD}. 
The energy of the $K = $ 0 (1) peak in fig. \ref{mg-24-IVD} 
is 20.2 (28.9) MeV, respectively.
The corresponding value obtained by using the QRPA by Peru et al. 
is 19.35 (27.44) MeV~\cite{Peru08}. 
The present TDHFB result of the $K =$ 0 (1) peak energy is higher by 0.9 (1.5) MeV 
than the QRPA result, respectively. 

From the isovector dipole strength functions, 
we calculated the photoabsorption cross section $\sigma(E)$ 
based on the strong coupling scheme~\cite{LinRes,BM}
\begin{widetext}
\begin{eqnarray}
  \sigma(E) &=& \frac{4 \pi^2 e^2 E}{c \hbar} \sum_{K = -1}^{1} \frac{dB(E, f_{1 K})}{d E}\, , \\
  \frac{dB(E, f_{1 K})}{d E} &=& - \frac{1}{\pi \hbar \varepsilon} 
   {\rm Im} \int_{0}^{\infty} dt\,  \langle \Psi (t) | f_{1 K} |\Psi (t) \rangle 
   e^{ i E t / \hbar - \gamma t }\, ,  \nonumber \\
 & &     
\end{eqnarray}
\end{widetext}
where $\frac{dB(E, f_{1 K})}{d E}$ is a transition strength with isovector dipole operators 
$f_{1 K} = \frac{N}{A} \sum_{i = 1}^{Z} \left( r Y_{1 K} \right)_i$ for protons and  
$f_{1 K} = - \frac{Z}{A} \sum_{i = 1}^{N} \left( r Y_{1 K} \right)_i$ for neutrons 
with $K = \pm 1, 0$. 
The nuclear state $| \Psi (t) \rangle$ is the TDHFB state.
$\gamma$ is an artificial width used in the strength function and $\varepsilon$ is 
the small parameter used in the initial conditions (\ref{ini-1}) and (\ref{ini-2})
of the TDHFB equations.

In the right panel in fig. \ref{mg-24-IVD} we compare the photoabsorption cross section 
with the experimental results. 
The $K = $ 0 peak comes near the one of the experimental cross section, 
whereas the $|K| = 1$ peak is located in the higher energy region than 
the distribution of the peaks in the experimental data~\cite{Varlamov}.

The difference of the TDHFB results presented above in this section 
from the RPA results and experimental data might come partly 
from the restricted space of the basis functions 
($n_x + n_y + n_z \leq 5$) and the omission of the Coulomb force in the 
present TDHFB calculations. 
It is necessary to study the effects of the size of the space of basis functions 
and Coulomb force on the TDHFB results.

\section{Summary and concluding remarks} \label{sec:5}
In this article, we presented a practical numerical method of integrating the TDHFB equations 
with Gogny interaction. The three-dimensional harmonic oscillator eigen functions were used 
as the numerical basis functions. 
The conservation property of the energy, particle numbers, and center-of-mass of the 
method was illustrated. 
The method was applied to several typical spherical nuclei as well as axially deformed ones   
and the strength functions were obtained in the framework of the linear response theory. 
It was shown that the tendency of the energies of the low-energy peaks in the present TDHFB calculations  
was similar to the one of the experimental results, 
though the energies obtained in the TDHFB calculations tend to be lower than the QRPA and experimental 
results. On the other hand, the energies of the typical peaks in the energy region 
of the giant resonances tend to be larger than the QRPA and experimental results. 
We will study carefully how the numerical results depend on the space of basis functions 
and Coulomb force.

The present calculations were carried out under the conditions: 
i) The maximum major shell quantum number 
of the space of the harmonic oscillator basis functions 
was restricted to be four or five. 
ii) The Coulomb force was omitted in both HFB and TDHFB calculations. 
These restrictions were introduced to save the cpu time and 
check the feasibility of the present numerical codes of TDHFB. 

It is necessary to use larger space of basis functions and to include 
the Coulomb force so that we can deal with the problems of heavier, unstable nuclei 
in terms of the present TDHFB method.
When the larger space of the basis functions is taken, 
we need efficient numerical methods to save the cpu time of the iterations 
in the full Gogny-TDHFB calculations with the Coulomb force. 
The acceleration of the calculations of the Coulomb part 
might be realized by introducing a set of small number of {\it optimal} integration points 
and weights in the calculations of the Coulomb mean-field matrix elements.  
The new numerical codes by which we could save the cpu time are now being developed.

The author thanks Professors K.~Yabana and T.~Nakatsukasa for discussions and comments. 
He thanks Professor K.~Matsuyanagi for his warm encouragements. 
The author is grateful to Professor J.~A.~Maruhn for his suggestions and comments. 
The author is also grateful to Dr. S.~Ebata for discussions on the numerical methods. 
Part of the numerical calculations was carried out on SR16000 at YITP 
in Kyoto University.  

\end{document}